\documentclass[12pt]{article}

\def\be{\begin{equation}}
\def\ee{\end{equation}}

\newtheorem{theorem}{Theorem}
\newtheorem{lemma}[theorem]{Lemma}

\newtheorem{definition}[theorem]{Definition}
\newtheorem{proposition}[theorem]{Proposition}

\begin{document}

\title{Pseudodifferential operators \\ on ultrametric spaces \\ and ultrametric wavelets}

\author{A.Yu.Khrennikov, S.V.Kozyrev}

\maketitle

\begin{abstract}
A family of orthonormal bases, the ultrametric wavelet bases, is
introduced in quadratically integrable complex valued functions
spaces for a wide family of ultrametric spaces.

A general family of pseudodifferential operators, acting on
complex valued functions on these ultrametric spaces is
introduced. We show that these operators are diagonal in the
introduced ultrametric wavelet bases, and compute the
corresponding eigenvalues.

We introduce the ultrametric change of variable, which maps the
ultrametric spaces under consideration onto positive half--line,
and use this map to construct non--homogeneous generalizations of
wavelet bases.
\end{abstract}

\section{Introduction}

The present paper is devoted to investigation of
pseudodifferential operators on ultrametric spaces. Ultrametric
pseudodifferential operators were considered in
\cite{VVZ}--\cite{HuK}. The simplest example among these operators
is the Vladimirov $p$--adic fractional derivation operator, which
can be diagonalized by the $p$--adic Fourier transform (the field
of $p$--adic numbers is the example of ultrametric space). In the
present paper we introduce a wide family of pseudodifferential
operators on more general ultrametric spaces, which do not
necessarily possess a group structure. Since there is no Fourier
transform on general ultrametric space, the introduced
pseudodifferential operators cannot be diagonalized using this
method. Instead of this we introduce and apply the method of
ultrametric wavelets.

In \cite{wavelets} the basis of $p$--adic wavelets in the space
$L^2(Q_p)$ of quadratically integrable functions on the field of
$p$--adic numbers was introduced and it was proven that this basis
is a basis of eigenvectors for the Vladimirov operator. Also the
relation to the standard wavelet analysis on real line was
discussed. The wavelet analysis is a well established approach
used in a broad field of applications (see for instance the review
\cite{Daubechies2}).

In paper \cite{nhoper} a family of pseudodifferential operators in
the space $L^2(Q_p)$, diagonal in the basis of $p$--adic wavelets,
but not diagonalizable by the Fourier transform, was built, and
the corresponding eigenvalues were computed. In the present paper
we generalize the results of papers \cite{wavelets} and
\cite{nhoper} onto the case of a wide family of ultrametric
spaces.

We introduce bases of ultrametric wavelets on the considered wide
family of ultrametric spaces. These bases are analogous to
$p$--adic wavelet basis, constructed in \cite{wavelets}. We prove
that the ultrametric wavelet bases consist of eigenvectors for the
introduced pseudodifferential operators. For this class of
ultrametric spaces the wavelet analysis turns out to be an
effective substitute for the Fourier analysis.

The consideration of the present paper is motivated by
applications to ultrametric mathematical physics, for instance, to
replica approach and disordered phenomena. For $p$--adic
mathematical physics, see \cite{VVZ}, \cite{Vstring}--\cite{ABKO},
\cite{wavelets}, where the models of mathematical physics are
investigated with the help of $p$--adic analysis. For discussions
of the replica approach and applications of ultrametricity in
physics see for instance \cite{SpinGlass}, \cite{SpinGlass1}. In
\cite{ABK}, \cite{PaSu} it was shown that the Parisi matrix
describing the replica symmetry breaking (before the $n\to 0$
limit) is a discrete analogue of a $p$--adic pseudodifferential
operator. In \cite{Carlucci}, \cite{Carlucci1} the Parisi matrices
related to more general Abelian locally compact groups were
considered. In \cite{ABKO} the relation of ultrametric diffusion
and dynamics of macromolecules was discussed. Other applications
of $p$--adic mathematical physics are mathematical models in
biology and mental sciences \cite{Andr1}, \cite{Andr2}.

In the present paper we discuss the relation between the wavelet
analysis and the ultrametric analysis. This discussion follows the
line of research of \cite{wavelets}, where a $p$--adic wavelet
basis in the space of quadratically integrable functions with
$p$--adic argument was introduced. To be more specific, in
\cite{wavelets} it was shown that the natural map of $p$--adic
numbers onto positive real numbers (called the $p$--adic change of
variable) maps this basis onto the wavelet basis (in the space of
quadratically integrable functions on positive real half--line)
generated by the Haar wavelet (for $p$=2). For $p>2$ the $p$--adic
change of variable maps the basis of $p$--adic wavelets onto the
orthonormal basis in $L^2({\bf R_+})$ which is a simple
generalization of the basis generated by the Haar wavelet. The
basis vectors are complex valued compactly supported stepwise
functions, which take $p$ different values (the $p$--th complex
roots of 1).

In this paper we introduce specific maps for the considered
ultrametric spaces onto positive real numbers. The introduced maps
are surjective, they are one to one correspondences on the set of
full measure, and are continuous. We call these maps the
ultrametric changes of variable. They map ultrametric wavelet
bases onto some new orthonormal bases in $L^2({\bf R}_+)$. Note
that these bases are analogous to the wavelet basis generated by
the Haar wavelet. The main difference with the $p$--adic case is
that the image of an ultrametric wavelet basis contains the
vectors which are, up to shifts and dilations, the images of
$p$--adic wavelets with different $p$. We call the image of
ultrametric wavelet basis the non-homogeneous wavelet basis in
$L^2({\bf R}_+)$. The non--homogeneity here means that, unlike in
the case of the usual wavelet bases, vectors of non--homogeneous
wavelet basis can not be constructed using shifts and dilations of
fixed wavelet.

The structure of the present paper is as follows.

In Section 2 we construct a family of the ultrametric spaces under
consideration, and build measures on these spaces.

In Section 3 we introduce the orthonormal ultrametric wavelet
bases in the spaces of quadratically integrable functions on
spaces, considered in Section 2.

In Section 4 we introduce pseudodifferential operators, acting on
complex valued functions on spaces, defined in Section 2, and
prove that these pseudodifferential operators are diagonal in the
bases of ultrametric wavelets.

In Section 5 we introduce the ultrametric change of variable,
which maps the ultrametric spaces under consideration onto
positive half--line, and use this map to construct
non--homogeneous generalizations of wavelet bases.

\section{Construction of the ultrametric space}

In the present section we define a family of ultrametric spaces
related to trees. A tree is a graph without loops. For general
discussion of trees see \cite{Serre}.

An ultrametric space is a metric space with the metric $|xy|$ (the
distance between $x$ and $y$), which  satisfies the strong
triangle inequality
\[
|ab|\le\hbox{ max }(|ac|,|cd|),\qquad \forall c
\]

Consider an arbitrary tree (finite or infinite), such that the
path in the tree between arbitrary two vertices is finite, and the
number of edges incident to each of the vertices is finite. If a
vertex $I$ is incident to $p_I+1$ edges, we will say that the
branching index of $I$ is $p_I$. Examples of this kind of trees
are the Bruhat--Tits trees (when the branching index is constant).

The absolute of a tree will be an ultrametric space (with respect
to the naturally defined metric). Consider two equivalent
definitions of the absolute of the tree.

The first definition is as follows. The infinitely continued path
with the beginning in vertex $I$ is a path with the beginning in
$I$, which is not a subset of a larger path with the beginning in
$I$. The space of infinitely continued paths in the tree, which
begin in some vertex $R$ (that is, the root) is called the
absolute of the tree. Obviously the definition of the absolute of
the tree does not depend on the choice of $R$ (taking any other
vertex $A$ leads to an equivalent definition).

The equivalent definition of the absolute is as follows: the
absolute is the space of equivalence classes of infinitely
continued paths in the tree, such that any two paths in one
equivalence class coincide starting from some vertex (i.e. the
tails of the paths in one equivalence class are the same). If we
choose in each of the equivalence classes the paths, which begin
in vertex $R$, we will reproduce the first definition.

We consider trees with a partial order (or directed trees), where
the partial order is defined in the following way. Fix the vertex
$R$ and the point $\infty$ at the absolute. To fix the point
$\infty$ at the absolute means that have to fix the infinitely
continued path $R\infty$ from the vertex $R$ to $\infty$. The
point $\infty$ we will call the infinite point, or the infinity.
We define the following natural partial order on the set of
vertices of the tree: $J>I$ if $J$ belongs to the path $I\infty$.

We denote the absolute of the tree by $X$. Let us construct an
ultrametric and a measure on $X$.

For the points $x$, $y$ of the absolute there exists a unique path
$xy$ in the tree. The notation $xy$ should be understood in the
following way. Since the points $x$, $y$ of the absolute are
identified with the paths $Rx$ and $Ry$, the path $xy$ will be
contained in $Rx \bigcup Ry$. Then there exists a unique vertex
$A$ satisfying \be\label{A} Rx=RAx,\qquad Ry=RAy,\qquad Ax\bigcap
Ay=A \ee The notation $ABC$ means that $AC=AB\bigcup BC$. Then
\[
xy=Ax\bigcup Ay
\]

Consider the paths $x\infty$ and $y\infty$. There exists a unique
smallest (in the introduced partial order) vertex $I$ such that
\be\label{I} x\infty=xI\infty,\qquad y\infty=yI\infty \ee We have
\[
x\infty=xI\bigcup I\infty,\quad y\infty=yI\bigcup I\infty,\quad
xy=xI\bigcup I y
\]

We have three possibilities.

1) Let $I>R$. Then
\[
xy\bigcap R\infty=I
\]
Consider the (non--maximal) path $RI=I_0I_1\dots I_k$,
$R=I_0<I_1<\dots<I_k=I$.

Define the distance between $x$ and $y$ as the following product
of branching indices: \be\label{distance1} |xy|=\prod_{j=1}^{k}
p_{I_{j}} \ee

2) Let $I\le R$. Then the vertex $R$ lies at the path $I\infty$.
In this case we have the path $RI=I_0I_1\dots I_k$,
$R=I_0>I_1>\dots>I_k=I$.

Define the distance between $x$ and $y$ as follows:
\be\label{distance2} |xy|=\prod_{j=0}^{k-1} p_{I_{j}}^{-1} \ee

When $I=R$, the product above contains empty set of multipliers,
and we define the distance as $|xy|=1$.

3) Let $I$ and $R$ are incomparable. In this case there exists a
unique supremum $J$ in the sense of the introduced partial order
in the tree:
\[
J=\hbox{ sup }(I,R)
\]
i.e. $J$ is the smallest vertex larger than both $R$ and $I$:
\[
I\infty=IJ\infty,\quad R\infty=RJ\infty
\]
Consider the paths $RJ=J_0J_1\dots J_k$, $R=J_0<J_1<\dots<J_k=J$;
and $IJ=I_0I_1\dots I_l$, $I=I_0<I_1<\dots <I_l=J$
(correspondingly $J=J_k=I_l$).

Define the distance between $x$ and $y$ as follows:
\be\label{distance3} |xy|=\prod_{m=1}^{k} p_{J_{m}}
\prod_{n=1}^{l}p^{-1}_{I_{n}} \ee

Summing up the above three cases, the introduced distance between
$x$ and $y$ can be described as follows.

Put into correspondence to an edge in the tree the branching index
of the largest vertex of the edge (this definition is correct,
since any two vertices, connected by edge, are comparable).

For the points $x$ and $y$ of the absolute consider vertex $I$,
where the paths $x\infty$ and $y\infty$ merge. Then the distance
$|xy|$ is introduced as the product of branching indices of edges
in the directed path $RI$ in the degrees $\pm 1$, where branching
indices of increasing edges are taken in the degree $+1$, and
branching indices of decreasing edges are taken in the degree
$-1$. Here an edge is called increasing, if the end of the edge is
larger than the beginning, and is called decreasing in the
opposite case.

The following lemma can be proved by direct computation.

\begin{lemma}\label{isultrametric}
The function $|xy|$ is an ultrametric (i.e. it is nonnegative,
equal to zero only for $x=y$, symmetric, and satisfies the strong
triangle inequality):
$$
|xy|\le \hbox{ max }(|xz|,|yz|),\qquad \forall z
$$
\end{lemma}

\bigskip

\noindent{\it Proof}\qquad To prove that $|xy|$ is an ultrametric,
it is sufficient to prove that $|xy|$ satisfies the strong
triangle inequality (the other conditions, which are necessary for
ultrametricity are obvious).

Consider the points $x$, $y$, $z$ at the absolute and the
corresponding paths $x\infty$, $y\infty$, $z\infty$. Then the
paths $x\infty$, $y\infty$ coincide, starting from some vertex $I$
(we consider these paths as increasing paths to the point
$\infty$). Analogously, the paths $y\infty$, $z\infty$ coincide,
starting from some vertex $J$; the paths $x\infty$, $z\infty$
coincide, starting from some vertex $K$.

Since vertices $I$ and $K$ lie at the increasing path $x\infty$,
these vertices are comparable. Analogously, the vertices $I$ and
$J$ are comparable; as well as the vertices $J$ and $K$. Therefore
the set of vertices $I$, $J$, $K$ is an ordered set.

There are two possibilities: or $I=J=K$, or some of the vertices
do not coincide. If $I=J=K$, then by (\ref{distance1}),
(\ref{distance2}), (\ref{distance3})
$$
|xy|=|xz|=|yz|
$$

Let $I>J$. Then the increasing paths $y\infty$ and $z\infty$
coincide, starting from $J$, and coincide with the path $x\infty$,
starting from $I$, which implies that $I=K$. Therefore, by
(\ref{distance1}), (\ref{distance2}), (\ref{distance3})
$$
|xy|=|xz|>|yz|
$$
i.e. the strong triangle inequality is satisfied.

Analogously, with the other choice of the order on the set $I$,
$J$, $K$ we again will obtain the strong triangle inequality,
which finishes the proof of the lemma.

\bigskip

We have defined the ultrametric on the absolute of the tree. In
the topology corresponding to the defined ultrametric, the
absolute $X$ will be locally compact. For the Bruhat--Tits tree
the construction of ultrametric reduces exactly to the definition
of $p$--adic distance.

Define the measure $\mu$ on the absolute of the tree, which for
the case of the Bruhat--Tits tree will reduce to the Haar measure
on $p$--adic numbers. To define the measure $\mu$, it is enough to
define this measure on the disks $D_I$, where $D_I$ is the set of
all the infinitely continued paths incident to the vertex $I$
which intersect the path $I\infty$ only at the vertex $I$.

Define the diameter $d_I$ of the disk as the supremum of the
distance $|xy|$ between the paths $Ix$ and $Iy$ in $D_I$. Then
$D_I$ is the ball of radius $d_I$ with its center on any of $Ix\in
D_I$.

\begin{definition}\label{measureeqradius}
The measure $\mu(D_I)$ of the disk $D_I$ is equal to the disk
diameter.
\end{definition}

Since the disk $D_I$ contains $p_I$ maximal subdisks, which by
definitions of the ultrametric and the measure have the measure
$p_I^{-1}\mu(D_I)$, the measure $\mu$ is additive on disks. By
additivity  we can extend the measure on algebra generated by
disks ($\sigma$--additivity of the measure will follow from the
local compactness of the absolute, analogously to the case of the
Lebesgue measure). We denote $L^2(\mu, X)$ the space of the square
integrable (with respect to the defined measure) functions on the
absolute. Since the absolute $X$ is not a group, there is no
Fourier transform in $L^2(\mu, X)$. We are nevertheless able to
define the wavelet transform.

Define the enumeration on the set of directed edges (the edge is
directed, or has a direction, if we distinguish the beginning and
end of the edge). For each vertex $I$ in the tree we have $p_I+1$
edges incident to the vertex, $0\le p_I<\infty$. By definition
there exists a unique edge incident to the path $I\infty$.
Enumerate this edge by $-1$, and enumerate all the other $p_I$
edges by $x_I=0,\dots, p_I-1$ in an arbitrary way. Note that the
direction of the edge is important: since every edge has a
beginning and an end, it corresponds to two directed edges (with
the opposite direction) with two different numerations. We also
take all the edges at the path $R\infty$, directed from the
$\infty$ to $R$ be enumerated by 0 (and by $-1$ if the edges are
directed in the opposite way).

Define the following enumeration of the points of the absolute $X$
by sequences of indices. Consider the point $x$ of the absolute.
Consider the paths $Rx$ and $R\infty$. There exists a unique
vertex $I$ such that
\[
Rx=RIx,\qquad R\infty=RI\infty,\qquad x\infty=xI\bigcup I\infty
\]
It is obvious that $R$ and $I$ are comparable, i.e. we have two
possibilities: $I\le R$ or $I>R$.

1) Let $I>R$ and the distance in the tree between $I$ and $R$ (the
number of edges in the path $IR$) is $\gamma$. Then we enumerate
the vertices in the corresponding path
$Ix=I_{-\gamma}I_{-\gamma+1}\dots$. The sequence corresponding to
$x$ can be written:
\[
x=x_{I_{-\gamma}}x_{I_{-\gamma+1}}\dots
x_{I_{-1}},x_{I_{0}}x_{I_{1}}\dots
\]
Here $x_J$ are the numbers of the edges directed from the higher
to the lower vertex in the path (this means that there is no $-1$
indices here, all the indices are in the set $0,\dots,p_J-1$).

2) Let $I=R$. Then we enumerate the vertices in the path
$Rx=I_0I_1\dots$. The sequence corresponding to $x$ can be
written:
\[
x=0,x_{I_0}x_{I_1}\dots
\]
This enumeration is the analogue of the expansion of a $p$--adic
number into a series over the degrees of $p$ or of the expansion
of a real number into infinite decimal fraction. In both these
expansions, numbers (real or $p$--adic) are parameterized by
sequences of digits. This suggests to call the introduced
parameterization of the absolute the digital parameterization.

\bigskip

\noindent{\bf Remark}\qquad The defined above parameterization
allows to put in correspondence to the vertex $I$ the point of the
absolute with the enumeration $I 0\dots$, which we will denote by
the same symbol $I$.

\section{The wavelet basis in $L^2(\mu, X)$}

For the vertex $I$ of the tree, define the function $\Omega_I(x)$
on the absolute, which is equal to the characteristic function of
the disk $D_I$.

Define the ultrametric wavelet as the function $\psi_{Ij}(x)$ on
the absolute, where $I$ is the vertex of the tree and
$j=1,\dots,p_I-1$, given by the formula \be\label{wavelet}
\psi_{Ij}(x)={e^{2\pi i j x_I
p_{I}^{-1}}\Omega_I(x)\over\sqrt{\mu(D_I)}} \ee

The point $x$ of the absolute is a class of equivalence of
infinitely continued paths. In this class there exists a path
which begins in the vertex $I$. Then $\Omega_I(x)$ is equal to 0
or 1 depending on the direction of this path at $I$: $\Omega_I(x)$
is equal to 0 if the edge of the path $x$ at vertex $I$ is
directed onto infinity, and $\Omega_I(x)$ is equal to 0 otherwise.

Note that the definition of the wavelet depends on the enumeration
of the edges of the tree (but the supports of the wavelet do not
depend on the enumeration).

\begin{theorem}\label{basis}
$\{\psi_{Ij}\}$ is an orthonormal system of functions in $L^2(\mu,
X)$. If all the infinitely continued directed paths in the tree
are infinite, then $\{\psi_{Ij}\}$ is a basis in $L^2(\mu, X)$.
\end{theorem}

\noindent{\it Proof}\qquad Consider the scalar product
\begin{equation}\label{pairing}
\langle \psi_{Ij},\psi_{I'j'}\rangle=
{1\over\sqrt{\mu(D_I)\mu(D_{I'})}}\int e^{-2\pi i j x_I
p_{I}^{-1}}e^{2\pi i j' x_{I'} p_{I'}^{-1}}
\Omega_I(x)\Omega_{I'}(x) d\mu(x)
\end{equation}
The expression above can be non--zero only when $I\ge I'$ or $I\le I'$.
Without loss of generality we choose $I\le I'$. In this case
\[
\Omega_I(x)\Omega_{I'}(x)=\Omega_I(x)
\]

Consider $I<I'$. Then for the integral at the RHS of (\ref{pairing})
we get
\[
{1\over\sqrt{\mu(D_I)\mu(D_{I'})}}e^{2\pi i j' x_{I'}
p_{I'}^{-1}}\int e^{-2\pi i j x_I p_{I}^{-1}} \Omega_I(x)
d\mu(x)=0
\]
since $x_{I'}$ is a constant on $D_I$.

Therefore, the scalar product (\ref{pairing}) can be non--zero
only for $I=I'$, then we obtain for (\ref{pairing})
\[
\langle \psi_{Ij},\psi_{Ij'}\rangle= {1\over\mu(D_I)}\int e^{2\pi
i (j'-j) x_{I} p_{I}^{-1}} \Omega_I(x)d\mu(x) =\delta_{jj'}
\]
We get for (\ref{pairing})
\[
\langle \psi_{Ij},\psi_{I'j'}\rangle=\delta_{II'}\delta_{jj'}
\]
which proves that vectors $\psi_{Ij}$ are orthonormal.

To prove that if all the infinitely continued directed paths in
the tree are infinite the set of vectors $\{\psi_{Ij}\}$ is an
orthonormal basis (i.e. it is total in $L^2(\mu, X)$), we use the
Parsevale identity. Since the set of indicators (characteristic
functions) of the disks $D_I$ is total in $L^2(\mu,X)$, proving
that $\{\psi_{Ij}\}$ is a total system requires only to check the
Parsevale identity for the indicator $\Omega_I(x)$.

We have for the normed indicator the following scalar product:
\begin{equation}\label{completeness}
{1\over\sqrt{\mu(D_J)}}\langle\Omega_{J},\psi_{Ij}\rangle=
{1\over\sqrt{\mu(D_I)\mu(D_J)}}\langle\Omega_{J}(x),e^{2\pi i j
x_I p_{I}^{-1}}\Omega_I(x)\rangle
\end{equation}
which is equal to
\[
{\sqrt{\mu(D_J)\over\mu(D_I)}}e^{2\pi i j x_I p_{I}^{-1}}
\]
for $J< I$, and to zero otherwise.

This implies the following identity: \be\label{pars}
\sum_{Ij}\left|{1\over\sqrt{\mu(D_J)}}\langle\Omega_{J},\psi_{Ij}\rangle\right|^2=
\mu(D_J)\sum_{I> J;j}{1\over{\mu(D_I)}}=\mu(D_J)\sum_{I>
J}{p_I-1\over{\mu(D_I)}} \ee Consider the increasing sequence
$J\infty$, $J=I_0<I_1<\dots$ of vertices starting from $J$. We
will consider both the cases when this sequence is finite or
infinite (when the sequence is finite, we will denote the largest
vertex in this sequence by $I_f$; this vertex can be identified
with the infinite point $\infty$ of the absolute). Since $f$ is
the length of the sequence $J\infty$, for the case when the
sequence $J\infty$ is infinite, we will say that $f$ is infinite.

The following property is satisfied
\[
\mu(D_{I_k})=\mu(D_J)\prod_{l=1}^{k}p_{I_l}
\]
This implies for (\ref{pars}) the following
\[
\mu(D_J)\sum_{I>
J}{p_I-1\over{\mu(D_I)}}=\sum_{k=1}^{f}{p_{I_k}-1\over{\prod_{l=1}^{k}p_{I_l}}}=
\sum_{k=1}^{f}
\left[\left(\prod_{l=1}^{k-1}p_{I_l}\right)^{-1}-\left(\prod_{l=1}^{k}p_{I_l}\right)^{-1}\right]
\]
which is equal to
\[
1-\left(\prod_{l=1}^{f}p_{I_l}\right)^{-1}
\]
when $f$ is finite, and to
\[
\lim_{f\to\infty}\left[1-\left(\prod_{l=1}^{f}p_{I_l}\right)^{-1}\right]=1
\]
when $f$ is infinite.

It means that if all the infinitely continued directed paths in
the tree are infinite the Parsevale identity is satisfied, and
that $\{\psi_{Ij}\}$ is an orthonormal basis in $L^2(\mu,X)$, thus
proving the theorem.

\bigskip

We call this basis the basis of the ultrametric wavelets. For the
$p$--adic case this basis reduces to the basis of $p$--adic
wavelets introduced in \cite{wavelets}.

\section{Pseudodifferential operators}

In the present section we construct a family of ultrametric
pseudodifferential operators, which will be diagonal in the basis
of ultrametric wavelets.

Consider the operator in $L^2(\mu, X)$ \be\label{generator} T
f(x)=\int T(x,y)(f(x)-f(y))d\mu(y) \ee

Introduce some conditions on the kernel $T(x,y)$ of the operator
(\ref{generator}).

\begin{definition}\label{transprob} We consider the class of kernels $T(x,y)$,
which are nonnegative and depend (as a function of two variables
$x$ and $y$) only on the highest (in the sense of the partial
order in the tree) point $A(x,y)$ lying at the path between $x$
and $y$ in the tree.
\end{definition}

\begin{lemma}\label{01}
The function $T(x,y)$ is symmetric, positive and locally constant
with respect to $y$ for a fixed $x$ (outside any vicinity of $x$),
and, for an arbitrary fixed $x$, the following condition is
satisfied: \be\label{lemma1} T(x,y)=\hbox{ const }, \qquad \hbox{
if  }\quad |xy|=\hbox{ const } \ee
\end{lemma}

\begin{theorem}\label{02} The function of the form
\be\label{lemma3} T(x,y)=\sum_{I} T^{(I)} \delta_{|I|,|xy|}
\Omega_I(x) \ee where $T^{(I)}\ge 0$, satisfies the conditions of
lemma \ref{01}, and an arbitrary function satisfying
(\ref{lemma1}) can be represented in the form (\ref{lemma3}).
\end{theorem}

Here $|I|=\mu(D_I)$ is the diameter if the disk $D_I$, which
consists of the paths incident to the vertex $I$ and directed in
the opposite direction to the infinity (and the disk diameter is
equal to the measure of the disk). Remind that the function
$\Omega_I(x)$ is the characteristic function of the disk $D_I$.

\bigskip

\noindent{\it Proof}\qquad The positivity of $T(x,y)$ is obvious.

Let us prove the symmetricity of $T(x,y)$. We have \be\label{T-T}
T(x,y)-T(y,x)= \sum_{I} T^{(I)} \delta_{|I|,|xy|}
\left(\Omega_I(x)-\Omega_I(y)\right) \ee In order to prove that
this expression is equal identically to zero, consider the case
when $x$ is such that the following characteristic function is
non--zero: $\Omega_I(x)=1$. This implies \be\label{x-n} |xI|\le
|I| \ee If $\delta_{|I|,|xy|}\ne 0$, then \be\label{x-y} |xy|= |I|
\ee Formulas (\ref{x-n}), (\ref{x-y}) and ultrametricity of the
absolute imply that $\Omega_I(y)=1$. Therefore, the corresponding
terms in (\ref{T-T}) cancel. This proves that the $T(x,y)$ given
in (\ref{lemma3}) is symmetric. Let us prove now that it satisfies
(\ref{lemma1}).

Fix $x\in D_I$ at the absolute. Then for $y$ lying at the sphere
with the center $x$ and the radius $|I|$ we have
\[
\delta_{|I|,|xy|}=1
\]
Also $\Omega_I(x)$ is a constant on this sphere. Therefore
$T(x,y)$ will be a constant on the considered sphere and
(\ref{lemma1}) will be satisfied.

This proves that $T(x,y)$ satisfying the conditions of the present
theorem will satisfy lemma \ref{01}.

Vice versa, it is easy to see that the kernel (\ref{lemma3}) for
$x$, $y$ lying in the disk with the center $I$ and the radius
$|I|$, and satisfying $|xy|=|I|$, takes the value $T^{(I)}$.

Since all the space $x,y\in X\times X$ is the disjoint union of
such a subsets, therefore, taking an arbitrary positive $T^{(I)}$
we are able to construct an arbitrary kernel satisfying
(\ref{lemma1}). This finishes the proof of the theorem.

\begin{theorem}\label{03} Let the kernel (\ref{lemma1})
satisfies the condition of convergence of all the integrals in
(\ref{lambdagn}) for any $I$. Then the operator (\ref{generator})
is a selfadjoint (and moreover, positive) operator in $L^2(\mu,X)$
with a dense domain and the wavelets $\psi_{Ij}$ are eigenvectors
for the operator (\ref{generator}): \be\label{lemma2}
T\psi_{Ij}(x)=\lambda_I \psi_{Ij}(x) \ee with the eigenvalues
\be\label{lambdagn} \lambda_I= \int_{|Iy|>|I|}T(I,y)d\mu(y)+
T(I,I1)\mu(D_I) \ee
\end{theorem}

Here in the notations $|Iy|$, $T(I,y)$, $T(I,I1)$, symbol $I$ is
the point of the absolute, corresponding to vertex $I$ in the
sense of the remark at the end of Section 2. Vertex $I1$ is the
maximal vertex, which is less than $I$ and has the numeration,
obtained from the numeration of $I$ by adding of 1 (in $T(I,I1)$
we mean the corresponding points of the absolute).

\bigskip

\noindent{\it Proof}\qquad To prove the present theorem we use
Lemma \ref{01}. Consider the wavelet $\psi_{Ij}$. Then
\[
T\psi_{Ij}(x)=\int T(xy)\left(\psi_{Ij}(x)-\psi_{Ij}(y)
\right)d\mu(y)
\]

Consider the following cases.

1) Let $x$ lies outside $D_I$. Then Lemma \ref{01} implies
\[
T\psi_{Ij}(x)=-T(x,I)\int \psi_{Ij}(y)d\mu(y)=0
\]
Note that by (\ref{lemma1}) $T(x,I)$ does not depend on the
enumeration of the points of the absolute.

2) Let $x\in D_I$. Then again by Lemma \ref{01}
\[
T\psi_{Ij}(x)= \left(\int_{|xy|>|I|}+ \int_{|xy|=|I|}+
\int_{|xy|<|I|}\right)T(x,y) (\psi_{Ij}(x)-\psi_{Ij}(y))d\mu(y)=
\]
\[
= \left(\int_{|xy|>|I|}+ \int_{|xy|=|I|}\right)T(x,y)
(\psi_{Ij}(x)-\psi_{Ij}(y))d\mu(y)=
\]
\[
= \psi_{Ij}(x)\int_{|xy|>|I|} T(x,y)d\mu(y)+ \int_{|xy|=|I|}T(x,y)
(\psi_{Ij}(x)-\psi_{Ij}(y))d\mu(y)=
\]
\[
= \psi_{Ij}(x)\int_{|Iy|>|I|}T(I,y)d\mu(y)+
T(I,I1)\psi_{Ij}(x)\mu(D_I)p_I^{-1}\sum_{l=1}^{p_I-1}
\left(1-e^{2\pi i j l}\right)
\]
The last equality follows from Lemma \ref{01} and the local
constance of $\psi_{Ij}$.

Since for $j=1,\dots,p-1 ~\hbox{ mod }p$ we have
\[
\sum_{l=1}^{p-1} \left(1-e^{2\pi i p^{-1}j l}\right)=p
\]
we obtain
\[
T\psi_{Ij}(x)= \psi_{Ij}(x)\left(\int_{|Iy|>|I|}T(I,y)d\mu(y)+
T(I,I1)\mu(D_I)\right)
\]
which gives (\ref{lambdagn}). Therefore the operator $T$ is well
defined on the basis in $L^2(\mu,X)$. Moreover, the obtained
eigenvalues are nonnegative. This finishes the proof of the
theorem.

\bigskip

The next proposition gives a simple representation for the
eigenvalues of the operator (\ref{generator}) with the kernel
(\ref{lemma3}).

\begin{proposition}\label{04} Let the following series
converge: \be\label{seriesconverge} \sum_{J>R} T^{(J)} \mu(D_J)
<\infty \ee Then the operator (\ref{generator}) corresponding to
the kernel (\ref{lemma3}) is a selfadjoint (and moreover,
positive) operator in $L^2(\mu,X)$, which is diagonal in the basis
of ultrametric wavelets and has the following eigenvalues in this
basis: \be\label{lemma4} \lambda_{I}= T^{(I)} \mu(D_I)+\sum_{J>I}
T^{(J)} \mu(D_J)(1-p_J^{-1}) \ee
\end{proposition}

Note that condition of convergence of the series
(\ref{seriesconverge}) is equivalent to convergence of the
integrals (\ref{lambdagn}).

\bigskip

\noindent{\it Proof}\qquad
Substituting (\ref{lemma3}) into (\ref{lambdagn}) we get
\[
\lambda_{I}= \int_{|Iy|>|I|}\sum_{J} T^{(J)} \delta_{|J|,|Iy|}
\Omega_J(I)d\mu(y)+ \sum_{J} T^{(J)} \delta_{|J|,|I,I1|}
\Omega_J(I)\mu(D_I)= \] \[ =\sum_{J>I} T^{(J)}
\mu(D_J)(1-p_J^{-1})+ T^{(I)} \mu(D_I)
\]
if the corresponding series converge.

Here we use the property
\[
\delta_{|J|,|I,I1|} \Omega_J(I)=\delta_{IJ}
\]

Since every two paths in the tree, which go to infinity, coincide
starting from some vertex, condition (\ref{seriesconverge}) is
equivalent to
\[
\sum_{J>I} T^{(J)} \mu(D_J) <\infty,\qquad \forall I
\]

This finishes the proof of the proposition.

\section{Relation to wavelets on real line}

In \cite{wavelets} the relation between the basis $\{\psi_{\gamma
j n}\}$ of $p$--adic wavelets and the basis of wavelets in the
space of quadratically integrable functions $L^2({\bf R}_+)$ on
positive half--line was discussed. The basis $\{\psi_{\gamma j
n}\}$ was called the basis of $p$--adic wavelets, since after the
natural map of $p$--adic numbers onto positive real numbers
(called the $p$--adic change of variable) this basis maps onto the
wavelet basis (in the space of functions on positive real
half--line) generated by the Haar wavelet (for $p$=2). For $p>2$
the $p$--adic change of variable maps the basis of $p$--adic
wavelets onto the orthonormal basis in $L^2({\bf R_+})$, which is
a simple generalization of the basis generated by the Haar
wavelet: the vectors of this basis are complex valued compactly
supported stepwise functions, which take $p$ different values
equal to the $p$--th complex roots of 1.

The wavelet basis in $L^2({\bf R}_+)$ is a basis given by shifts
and dilations of the mother wavelet function, cf.
\cite{Daubechies2}. The simplest example of such a function is the
Haar wavelet
\begin{equation}\label{haar}
\Psi(x)=\chi_{[0,\frac{1}{2}]}(x)- \chi_{[\frac{1}{2},1]}(x)
\end{equation}
(i.e, the difference of two characteristic functions).

The wavelet basis in $L^2({\bf R})$ (or basis of multiresolution
wavelets) is the basis
\begin{equation}\label{basis1}
\Psi_{\gamma  n}(x)=2^{-{\gamma\over 2}}\Psi(2^{-\gamma}x-n),\quad
\gamma\in {\bf Z},\quad n\in {\bf Z}
\end{equation}

The $p$--adic wavelets $\psi_{\gamma j n}(x)$, see
\cite{wavelets}, are defined in the way similar to the definition
of $\psi_{Ij}(x)$ in the present paper:
\[
\psi_{\gamma j n}(x)=p^{-{\gamma\over 2}} \chi(p^{\gamma-1}j x)
\Omega(|p^{\gamma} x-n|_p);\quad \gamma\in {\bf Z}, n\in Q_p/Z_p,
j=1,\dots,p-1
\]
The following $p$--adic change of variable was considered:
\[
\eta:Q_p \to {\bf R}_+
\]
\begin{equation}\label{change}
\eta:\sum_{i=\gamma}^{\infty} a_i p^{i} \mapsto
\sum_{i=\gamma}^{\infty} a_i p^{-i-1},\quad a_i=0,\dots,p-1,\quad
\gamma \in {\bf Z}
\end{equation}
which maps the wavelet basis on the basis of $p$--adic wavelets.
The following theorem was proven:

\bigskip

\noindent {\bf Theorem}.\quad {\sl For $p=2$ the map $\eta$,
defined by (\ref{change}), maps the orthonormal basis of wavelets
in $L^2({\bf R}_+)$ (generated from the Haar wavelet) onto the
basis of eigenvectors of the Vladimirov operator ($p$--adic
wavelets):
\begin{equation}\label{mapofbasis}
\eta^*:\Psi_{\gamma \rho(n)}(x)\mapsto (-1)^{n} \psi_{\gamma 1
n}(x)
\end{equation}
}

\bigskip

For general $p$ the $p$--adic change of variable, applied to the
basis of $p$--adic wavelets, will generate a basis in $L^2({\bf
R}_+)$ of vectors which, up to multiplication by numbers, have the
form
$$
\Psi^{(p)}_{\gamma  n}(x)=p^{-{\gamma\over
2}}\Psi^{(p)}(p^{-\gamma}x-n),\quad \gamma\in {\bf Z},\quad n\in
{\bf Z}_{+}
$$
where ${\bf Z}_{+}$ is the set of positive integers and
$$
\Psi^{(p)}(x)=\sum_{l=0}^{p-1}e^{2\pi i l
p^{-1}}\chi_{[lp^{-1},(l+1)p^{-1}]}(x)
$$
This basis is a generalization of the basis of wavelets, generated
by the Haar wavelet (and can be extended into a basis in $L^2({\bf
R}_+)$, if we take $n \in {\bf Z}_{+}$).

Constructed in the present paper basis $\{\psi_{Ij}\}$ gives rise
to a new basis in $L^2({\bf R}_+)$, which is a generalization of
the wavelet basis.

In the present paper we build a generalization of the map $\eta$,
which we will call the ultrametric change of variable and denote
by $\rho:X\to {\bf R}_+$. For the point $x$ at the absolute
\[
x=x_{I_{\gamma}}x_{I_{\gamma+1}}\dots
x_{I_{-1}},x_{I_{0}}x_{I_{1}}\dots;\qquad x_I=0,\dots, p_I-1,\quad
\gamma \in {\bf Z}
\]
the map $\rho$ looks as follows
\begin{equation}\label{newchange}
\rho: x\mapsto \sum_{k=\gamma}^{-1}x_{I_{k}}
\prod_{l=k}^{-1}p_{I_l}+
\sum_{k=0}^{\infty}x_{I_{k}}\prod_{l=0}^{k}p_{I_l}^{-1}
\end{equation}
for negative $\gamma$ and
\begin{equation}\label{newchange1}
\rho:x\mapsto
\sum_{k=\gamma}^{\infty}x_{I_{k}}\prod_{l=0}^{k}p_{I_l}^{-1}
\end{equation}
for positive $\gamma$.

This map is not a one--to--one map (but it is a one--to--one map
almost everywhere). The map $\rho$ is continuous and moreover, one
can prove the following lemma:

\begin{lemma}\label{l3}
The map $\rho$ satisfies the H\"older inequality
\begin{equation}\label{norm}
|\rho(x)-\rho(y)| \le |xy|
\end{equation}
\end{lemma}

Note that $|\cdot|$ at the LHS of (\ref{norm}) is the modulus of
real argument and at the RHS is the ultrametric on the absolute
$X$.

\bigskip

\noindent{\it Proof}\qquad Consider
\[
x=x_{I_{\alpha}}x_{I_{\alpha+1}}\dots
x_{I_{-1}},x_{I_{0}}x_{I_{1}}\dots; \qquad
y=y_{J_{\beta}}y_{J_{\beta+1}}\dots
y_{J_{-1}},y_{J_{0}}y_{J_{1}}\dots.
\]
where we assume without loss of generality that $\alpha\le\beta$.
For simplicity we assume $0\le\alpha\le\beta$. In this case
\[
x=0,0\dots 0x_{I_{\alpha}}\dots; \qquad
\rho(x)=\sum_{k=\alpha}^{\infty}x_{I_{k}}\prod_{l=0}^{k}p_{I_l}^{-1}
\]
\[
y=0,0\dots 0y_{I_{\beta}}\dots; \qquad\rho(y)=
\sum_{k=\beta}^{\infty}y_{J_{k}}\prod_{l=0}^{k}p_{J_l}^{-1}
\]
Then
\[
|xy|=\prod_{l=0}^{\alpha-1}p_{I_l}^{-1}
\]

We have
\[
\rho(x)-\rho(y)=\sum_{k=\alpha}^{\beta-1}x_{I_{k}}\prod_{l=0}^{k}p_{I_l}^{-1}+\sum_{k=\beta}^{\infty}
\left[x_{I_{k}}\prod_{l=0}^{k}p_{I_l}^{-1}-y_{J_{k}}\prod_{l=0}^{k}p_{J_l}^{-1}\right]=
\]
\[
=|xy|\left(\sum_{k=\alpha}^{\beta-1}x_{I_{k}}\prod_{l=\alpha}^{k}p_{I_l}^{-1}+
\sum_{k=\beta}^{\infty}
\left[x_{I_{k}}\prod_{l=\alpha}^{k}p_{I_l}^{-1}-
y_{J_{k}}\prod_{l=\alpha}^{k}p_{J_l}^{-1}\right]\right)\le
\]
\[
\le
|xy|\left(\sum_{k=\alpha}^{\beta-1}\left(p_{I_{k}}-1\right)\prod_{l=\alpha}^{k}p_{I_l}^{-1}+
\sum_{k=\beta}^{\infty}
\left(p_{I_{k}}-1\right)\prod_{l=\alpha}^{k}p_{I_l}^{-1}\right)=
\]
\[=
|xy|\sum_{k=\alpha}^{\infty}\left(p_{I_{k}}-1\right)\prod_{l=\alpha}^{k}p_{I_l}^{-1}
=|xy|\lim_{f\to\infty}\left(1-\prod_{l=\alpha}^{f}p_{I_l}^{-1}\right)=|xy|
\]
which finishes the proof of the lemma.

\begin{lemma}\label{l4}
The map $\rho$ satisfies the conditions
\begin{equation}\label{in}
\rho: D_I\to \rho(I)+[0,\mu(D_I)]
\end{equation}
\begin{equation}\label{out}
\rho: X\backslash D_I\to {\bf
R}_+\backslash\{\rho(I)+[0,\mu(D_I)]\}
\end{equation}
up to a finite number of points.
\end{lemma}

Note that here we identify the vertex $I$ and the point at the
absolute with the enumeration $I0\dots$.

\bigskip

\noindent{\it Proof}\qquad For the vertex $I$, consider the points
\[
I=x_{I_{\alpha}}x_{I_{\alpha+1}}\dots
x_{I_{-1}},x_{I_{0}}x_{I_{1}}\dots x_{I_{\beta-1}}0\dots\] and
\[\tilde I=x_{I_{\alpha}}x_{I_{\alpha+1}}\dots
x_{I_{-1}},x_{I_{0}}x_{I_{1}}\dots
x_{I_{\beta-1}}p_{I_{\beta}}-1,\dots
\]
The first is the point at the absolute $X$ corresponding to the
vertex $I$, while the second is the first with the addition of the
tail of $p_{I_{\beta}}-1,\dots$.

\[
\rho(I)=\sum_{k=\alpha}^{-1}x_{I_{k}} \prod_{l=k}^{-1}p_{I_l}+
\sum_{k=0}^{\beta-1}x_{I_{k}}\prod_{l=0}^{k}p_{I_l}^{-1}
\]
\[
\rho(\tilde I)=\sum_{k=\alpha}^{-1}x_{I_{k}}
\prod_{l=k}^{-1}p_{I_l}+
\sum_{k=0}^{\beta-1}x_{I_{k}}\prod_{l=0}^{k}p_{I_l}^{-1}+
\sum_{k=\beta}^{\infty}(p_{I_{k}}-1)\prod_{l=0}^{k}p_{I_l}^{-1}
\]
We have
\[
\rho(\tilde
I)-\rho(I)=\sum_{k=\beta}^{\infty}(p_{I_{k}}-1)\prod_{l=0}^{k}p_{I_l}^{-1}=
\prod_{l=0}^{\beta-1}p_{I_l}^{-1}\sum_{k=\beta}^{\infty}(p_{I_{k}}-1)\prod_{l=\beta}^{k}p_{I_l}^{-1}=
\]
\[=
\mu(D_I)\lim_{f\to\infty}\left(1-\prod_{l=\beta}^{f}p_{I_l}^{-1}\right)=\mu(D_I)
\]
Then
\[
\rho(\tilde I)=\rho(I)+\mu(D_I)
\]
Using lemma \ref{l3}, we obtain the proof of the lemma.

\begin{lemma}\label{l5}
The map $\rho$ maps the measure $\mu$ on the absolute onto the
Lebesgue measure $l$ on ${\bf R}_+$: for any measurable subset
$S\subset X$ we have:
\[
\mu(S)=l(\rho(S))
\]
or in symbolic notations
\[
\rho:d\mu(x)\mapsto dx
\]
\end{lemma}

\noindent{\it Proof}\qquad Lemma \ref{l4} implies that disks in
$X$ map onto closed intervals   in ${\bf R}_+$ with conservation
of measure. The map $\rho:X\to {\bf R}_+$ is surjective, and since
nonintersecting disks map onto intervals that do not intersect or
have intersection of the measure zero (by lemma \ref{l4}), this
proves the lemma.

\bigskip

Therefore the corresponding conjugated map
\[
\rho^*:L^2({\bf R}_+)\to L^2(X)
\]
\begin{equation}\label{rhostar}
\rho^*f(x)=f(\rho(x))
\end{equation}
is a unitary operator. The inverse to this map will map basis of
ultrametric wavelets on the absolute $X$ on some basis of
functions on real positive half--line.

This and Lemmas \ref{l4}, \ref{l5} suggest the following
definition:

\begin{definition}\label{ultrawavelets}
We call the basis $\{\Psi_{Ij}\}$ in $L^2({\bf R}_{+})$, where
$\Psi_{Ij}=\rho^{-1*}\psi_{Ij}$, the basis of nonhomogeneous
wavelets on positive real half--line.
\end{definition}

The map $\rho^{-1*}$ between spaces of quadratically integrable
functions is well defined since the map $\rho$ is one to one on
the set of a complete measure.

We see that using the ultrametric change of variable $\rho$ we can
define the new examples of wavelets in $L^2({\bf R}_{+})$. The
name nonhomogeneous wavelets means that the basis
$\rho^{-1*}\{\psi_{Ij}\}$ in $L^2({\bf R}_{+})$ lacks translation
invariance (the shift of the wavelet is not necessarily a wavelet,
while for usual multiresolution wavelets this would be true).

The basis of nonhomogeneous wavelets combines wavelets corresponding to
$p$--adic wavelets with different $p$.

\bigskip

\noindent{\bf Acknowledgements}

The authors would like to thank I.V.Volovich, G.Parisi and
V.A.Avetisov for fruitful discussions and valuable comments.  One
of the authors (A.Kh.) would like to thank S.Albeverio for
fruitful discussions and support of $p$--adic investigations. This
paper has been partly supported by EU-Network ''Quantum
Probability and Applications''. One of the authors (S.K.) has been
partly supported by the Dynasty Foundation, CRDF (grant
UM1--2421--KV--02), The Russian Foundation for Basic Research
(project 02--01--01084), by the grant of the President of Russian
Federation for the support of scientific schools NSh 1542.2003.1,
and by the grant of The Swedish Royal Academy of Sciences on
collaboration with scientists of former Soviet Union.

\end{document}